\documentclass{ws-p9-75x6-50}

\begin{document}

\title{Nucleosynthesis in Advective Accretion Disk around Compact Object }

\author{Banibrata Mukhopadhyay}

\address{S.N.Bose National Centre for Basic Sciences, Block-JD,
Sector-III, Salt Lake, Calcutta-700098, India\\ E-mail: bm@boson.bose.res.in
\footnote{Presently working in: Physical 
Research Laboratory, Navrangpura, Ahmedabad-380009, India, E-mail: bm@prl.ernet.in}
}




\maketitle

\abstracts{
We study the nucleosynthesis in accretion disk around compact object.
When matter falls towards the compact
object, due to interactions in nuclear species nuclear energy is
released in the high temperature. In some region of
parameter space nuclear energy becomes comparable or higher to the
energy release due to the viscous effect of the disk. In those
regions disk may become unstable. Also the changes of nuclear
abundance in the disk are significant. 
}

In the accretion disk as matter comes closer to the compact object
its potential energy is converted to kinetic energy and then
to thermal energy. It is found observationally\cite{re} that incoming 
matter has the potential to become as hot as its virial 
temperature is of the order $10^{13}K$. By the inverse-Comptonization
and bremsstrahlung cooling effect the incoming
matter is usually cooled down to produce hard and soft states. 
However, the temperature of the sub-Keplerian region of the disk
is of the order $10^{9}K$ which is much higher compared to the central
temperature of star. After Big-Bang nucleosynthesis this is most
favourable temperature to occur nuclear reactions. So we were 
motivated to study the nucleosynthesis here. 

We follow Chakrabarti\cite{gut} to compute the basic properties
of the disk i.e., velocity, temperature, density etc. of the
incoming matter at different radii. Unlike
the stellar cases this high temperature is favourable to start 
proton-capture reactions and dissociation of elements like deuterium,
isotopes of helium etc. As a result either nuclear energy is released 
in the disk or absorbed from it. This effect of nuclear energy,
if significant enough with respect to the viscous and other
dissipation may influence on the structure of disk. The typical
density of the geometrically thin and optically thick accretion
disk is of the order $10^{-6}gm/cc$ for $10M_\odot$ black hole. If the massive
black hole is chosen the density becomes decrease more and more.
On the other hand in case of accretion disk around neutron star
with low magnetic field density could be higher with respect to 
that around black hole because the mass of neutron star could be
lower than black hole. So the nucleosynthesis strictly depends 
on mass of central 
compact object as reactions are temperature and density sensitive.
There is another basic difference between accretion disk around black 
hole and neutron star. In case of black hole accretion disk, close
to the black hole, attaining supersonic speed matter falls into it.
In case of neutron star, matter has to stop at
the hard surface and becomes subsonic close to central object.
As a result the density is higher close to the compact object 
in case of neutron star accretion disk and the nuclear reactions
are dominant. 

Away from the compact object, at temperature of the order $0.1-0.8\times
10^9K$ (i.e., $0.1-0.8T_9$, $T_9$ is the temperature in unit of $10^9K$)
the proton capture reactions are main source of nuclear energy.
In this process, energy is released via $^{18}\!O(p,\alpha)^{15}\!N
(p,\alpha)^{12}\!C$, $^6\!Li(p,^{3}\!He)^4\!He$, $^7\!Li(p,\alpha)^4\!He$,
$^{11}\!B(p,\gamma)3^{4}\!He$, $^{17}\!O(p,\alpha)^{14}\!N$ etc.\cite{mc} 
Closer to the compact object (say at $70-100r_g$ for
Shakura-Sunyaev viscosity parameter $\alpha=0.05-0.001$\cite{mc}, $r_g$ is 
Schwarzschild radius) when the temperature is at least
$0.8T_9$ dissociation of different elements via endothermic reactions
like $D(\gamma,n)p$, $^3\!He(\gamma,p)D$, $^4\!He(\gamma,D)D$ etc.\cite{mc} 
take place. In very closer to the central object where temperature
might be very high say $3-10T_9$ (depending on the $\alpha$ of the disk\cite{mc})
$^{12}\!C$, $^{16}\!O$, $^{24}\!Mg$ and $^{28}\!Si$ (if exist) are all 
destroyed completely. Some new species like $^{30}\!Si$, $^{46}\!Ti$, 
$^{50}\!Cr$ etc. are formed closer to the black hole or neutron star.
It is interesting to note that lithium can not be formed in the accretion
disk. Before the production of significant
lithium by spallation in the hot disk $^4\!He$ itself is 
dissociated completely\cite{mc}. All the above 
mentioned reactions and corresponding nuclear energy
release or absorption should take place in earlier radii in case of accretion
disk around neutron star with respect to that around black hole because
required temperature and density for nuclear reactions are greater
because of reason mentioned in second paragraph.   
In the certain region of parameter space where the temperature
of the disk is very very high (say at $T_9\ge 10$), it is seen that
due to high rate of production of nuclear energy the {\it real} sonic points may
be disappeared\cite{stab}. In those cases disk might be unstable
and incoming matter which was supposed to pass through the 
sonic point will not get any stable branch to proceed.
 
Another important feature which has been simulated in accretion disk
is the neutron torus\cite{mc}. In very high temperature
region enormous neutron is produced in the disk mainly by the dissociation
process of deuterium (sometimes helium is dissociated into deuterium and then
into neutron and proton). If the accretion rate is very low (say $0.01$ 
unit of Eddington rate) and viscosity is due to stochastic magnetic 
field, the neutron distribution forms a torus-like structure
called neutron torus in accretion disk. Being charged
neutral neutrons are not been affected by the viscosity (which is magnetic)
and are circled around central compact object after production. Here
neutron can be produced as upto $10\%$ of the total abundance
where initial neutron abundance was zero (we chose solar abundance 
as initial). Thus there is a possibility of formation 
of neutron rich element.

All the phenomena described above are the properties of advective accretion
disk where the cooling might be sufficient or insufficient. When the cooling
is insufficient high neutron abundance is very clear feature. Also in case
of moderate cooling, disk is hot enough to produce significant nuclear
energy and as a result the abundance of incoming matter must be changed
significantly. The outflow produced in the disk could carry out the new
isotopes and isotopes with changed abundance and may contaminate into
the surrounding. From the degree of contamination one can pinpoint about
inflow parameters.

\end{document}